# Ranking Swing Voters in Congressional Elections

## Steven Ambadjes

New York University, Courant Institute of Mathematical Sciences

251 Mercer Street

New York, NY 10012

January, 2014

\_\_\_\_\_\_\_\_\_\_\_\_\_\_\_\_\_\_\_\_\_\_\_\_\_\_\_\_\_\_

Micaela Fedele

\_\_\_\_\_\_\_\_\_\_\_\_\_\_\_\_\_\_\_\_\_\_\_\_\_\_\_\_\_\_

Tom LaGatta

# Ranking Swing Voters in Congressional Elections

**Abstract**

We present a model for quantitatively identifying swing voters in congressional elections. This is achieved by predicting an individual voter's likelihood to vote and an individual voter's likelihood to vote for a given party, if he votes. We make a rough prediction of these values. We then update these predictions by incorporating information on a municipality wide basis via aggregate data to enhance our estimate under the assumption that nearby voters have similar behavior, which could be due to social interaction or common external factors. Finally, we use a ranking scheme on these predictions to identify two key types of voter: 1) Voters who are likely to vote that we can convince to vote for a given party; and, 2) Voters who are likely to vote for a given party, if they vote, that we can convince to actually turn out to vote. Once these voters have been identified, a political campaign can use this information to micro-target voters and win more votes.

**Contents**

# 1. Introduction

Candidates in a given election employ a multitude of tactics to win votes. Whether they use methods such as polling or direct phone calls to potential voters, these candidates try to discern a voter's voting preferences and lock in as many potential votes as possible. While it is important to pay attention to all voters that will likely support you, emphasis should be placed on the swing voters in the election, since it is generally the swing voters that determine the outcome of an election (Gelman, Silver, Edlin, 2012). Since a campaign's objective is to maximize votes in a given election, this becomes a problem of identifying and ranking swing voters.

Identifying and ranking swing voters involves determining as accurately as possible the voting preferences for each voter. The importance of swing voters has long been known in political campaigns, and the emergence of Big Data allows ever more accurate predictions of a voter's voting preferences. This allows for a transition from a qualitative judgment to a more quantitative prediction of how a voter may vote. President Obama's Project Narwhal was a major appearance of Big Data applied to voter turn out. Although Project Narwhal in its entirety is beyond scope of this thesis, the underlying ideas can be implemented on a smaller scale and still be useful. As discussed in a recent Wired (Salmon, 2014) article, "It's increasingly clear that for smart organizations, living by numbers alone simply won't work. That's why they arrive at stage four: synthesis – the practice of marrying quantitative insight with old-fashioned subjective experience." That is precisely our intention in this thesis: use mathematical principles to make quantitative predictions, then enhance them via Bayesian inference with intuition discerned from empirical results.

Our objective in this thesis is to identify the most persuadable swing voters in Congressional campaigns and we will achieve this by ranking these voters. Part of doing this will involve trying to discern the party preference for unaffiliated voters. As we will discuss later, people behave differently depending on if its a Congressional, Presidential or off-year election. In this thesis, we will focus on Congressional campaigns and specifically use data from the 5th Congressional District of Connecticut to develop our model. To achieve our objective, we need to quantitatively describe the preferences of our voters. This is done by using the data set to assign them probability measures on likelihood to vote and likelihood to vote for a given party if they vote. We develop a model that makes as precise of a prediction as possible given the data for these values. This model is chiefly influenced by rational choice theory, statistical mechanics applied to socioeconomic phenomena, and Bayes' Rule. It should be noted that our final aim is not to predict the actual probabilities of each voter but, instead, probabilities that make sense on a relative basis to one another so we can implement an effective ranking scheme. Mayer-Schönberger and Cukier emphasize this point on their treatment of Big Data: "What we lose in accuracy at the micro level, we gain in insight at the macro level" (Mayer-Schönberger, Cukier 14). It is the intuition we get from the numbers that is significant, not the numbers themselves.

In the end, this is a practical project, with an aim to get real, usable results. To develop the intuition and as a proof of concept, the model was implemented in Microsoft Excel. The implementation of the model is not overly sophisticated, but the math that underlies the model is non-trivial. As discussed in Myerson's (2005) coverage of economic modeling, Microsoft Excel is a powerful tool in the implementation of statistical models.

This thesis is organized as follows. In chapter 2, we describe the data used underlying the model. In chapter 3, we explain the development of the model. In chapter 4, we discuss the ranking scheme used to identify swing voters. In chapter 5, we present the end results of the model and their implications. In chapter 6, we touch upon the theoretical underpinnings of our model. Finally, chapter 7 covers concluding thoughts and areas that model can be improved upon.

## 2. Data

The data used is drawn from the Connecticut Voter File, which is a list, made available by the State of Connecticut, that contains data about each registered voter in the state. In our model, we used the 5$^{th}$ Congressional District portion of the Connecticut voter file. There are 396,769 total voters spread over 41 towns in this district. In this data set, each voter is identified by their full name, which we have replaced with a voter id tag. Each voter has demographic information including city, zip, sex and date of birth. The party affiliation of each voter is given, and can be either Democrat (DEM), Republican (REP), or unaffiliated (UNA). The voting history (whether or not they've voted) for each year's general election from 2002 to 2012 is given for each voter. 2012, 2008 and 2004 elections were Presidential. 2010, 2006, and 2002 elections were Congressional. 2011, 2009, 2007, and 2005 elections were off-year. A random sample of 5 voters from the data set is shown below:

| VOTER_ID | CITY | ZIP | SEX | DOB | PARTY | 2012 | 2011 | 2010 | ... | 2006 | 2005 | 2004 |
|---|---|---|---|---|---|---|---|---|---|---|---|---|
| 3482 | Danbury | 06810 | F | 1972-11-21 | UNA | YES | | YES | | YES | | YES |
| 328203 | Middlebury | 06762 | M | 1958-03-04 | REP | | | YES | | | | YES |
| 47203 | Meriden | 06450 | F | 1986-04-23 | DEM | YES | | | | | | |
| 173812 | Litchfield | 06759 | M | 1955-09-01 | UNA | YES | YES | YES | | YES | | YES |
| 275024 | New Britain | 06052 | M | 1956-03-27 | UNA | YES | | | | | YES | |
| 94024 | Waterbury | 06708 | F | 1991-11-24 | UNA | YES | | YES | | | | |

District-wide, 54.7% of registered voters are female and 45.6% are male. The voters are split between parties as follows: 126,768 (31.95%) Democrats, 93,795 (23.64%) Republicans, and 176,206 (44.41%) unaffiliated. The breakdown is heterogeneous amongst the different municipalities. The difference in the number of Republicans compared to Democrats in these towns can be as small as 1-2% or as large as 10-15%. Most towns have more unaffiliated voters than any other type of voters, but the proportion of unaffiliated voters ranges from roughly 35% to 50%. For example, voters in Waterbury are 49% Democrat, 12% Republican and 39% unaffiliated, but voters in Bantam are much more evenly distributed with 25% Democrats, 23% Republicans, and 51% unaffiliated. Average voter turnout was 71% for Presidential elections, 51% for Congressional elections, and 38% for off-year elections.

## 3. The Model

The objective of our model is to identify swing voters in the 5$^{th}$ Congressional district of Connecticut in the next Congressional race, which will be held in 2014. We identify these swing voters via a ranking scheme. To do this, we need an estimate of an individual's probability to vote and probability to vote for a given party, if he votes. Throughout the model, we arbitrarily assume the objective is to win votes for the Democratic candidate and all the probabilities will be expressed in terms of predicting Democratic turnout. We will refer to individual i's probability to vote in 2014 as $p_{i,2014}$(Vote) and individual i's probability to vote Democrat in 2014, if he votes, as $p_{i,2014}$(Vote D | Vote). Note that the same analysis holds if the Democratic candidate was replaced with a Republican candidate. In this case, the values can either be calculated directly from the data or derived from the $p_{i,2014}$(Vote D | Vote) by noting that $p_{i,2014}$(Vote D | Vote) + $p_{i,2014}$(Vote R | Vote) = 1. That is, our model assumes if a voter votes he will either voter Democrat or Republican, so these probabilities must sum to one. [1] To obtain these values, we construct a model chiefly influenced by rational choice theory, socioeconomic application of statistics mechanics, and Bayesian inference. Abstractly, our model is rather

---

[1] Likewise, it is important to note that since the probability to vote and probability not to vote must sum to one, we have: $p_{i,2014}$(Abstain) = 1 - $p_{i,2014}$(Vote)

straightforward. We use the data to make a rough guess of each individual's prior probability. Once calculated, we will aggregate these probabilities on a municipality wide basis. Then, we use a principled update scheme on the individual priors to incorporate the information contained in the aggregates and obtain each individual's posterior probability estimates.

To calculate the prior probability to vote, we draw upon the voting history from our data set. As shown in the section above, the turnout averages are drastically different for each type of election (Presidential, Congressional, and off years). Since our model is concerned only with Congressional elections, we only consider results from Congressional years; that is, we are calculating $p_{i,2014}$(Vote | Congressional Year), which is the probability of voting, conditioned on it being a Congressional year. However, since we are only concerned with Congressional years, for ease of notation, we will refer to this simply as $p_{i,2014}$(Vote). We calculate this as follows:

$$p_{i,2014}(\text{Vote}) = (\text{\# of votes} + \varepsilon_v) / (\text{\# of eligible votes} + 1) \qquad \varepsilon_v \in (0,1)$$

where "# of votes" is the total number of times individual i voted in a Congressional election, "# of eligible votes" is the total number times individual i was eligible to vote in past Congressional elections. We consider a voter eligible to vote by using his year of birth to check if he is of legal age (18) to vote in a given election. Note that $p_{i,2014}$(Vote) is essentially the frequency that individual i voted in past Congressional elections; however, voting in none or all of the elections would lead to probabilities of 0 or 1, which are not well defined. Placing $\varepsilon_v$ in the numerator allows us to perturb the frequency such that the resulting values are valid probabilities between 0 and 1 (non-inclusive). Since priors are only a rough estimates to our probabilities, we choose $\varepsilon_v$ such that the model's predicted probabilities are reasonably close to the empirical results. [2]This yields $\varepsilon_v = .9$.

---

2  To calculate $\varepsilon_v$ and $\varepsilon_p$ we need to compare the model's predicted probabilities to empirical results. The empirical results are available on the Connecticut Secretary of the State's website (www.ct.gov/sots/). Due to limitations of the excel spreadsheet used to implement the model, we used a 10,000 sample

To calculate the prior probability to vote for a given party if a voter votes, we draw upon the voter's party affiliation from our data set. We assume that each voter can be registered either Democrat, denoted DEM, Republican, denoted REP, or unaffiliated, denoted UNA. If this voter votes, he votes either for the Republican candidate or the Democratic candidate. The probability voter i votes for a certain party is calculated as following:

$$P_{i,2014}(\text{Vote D} \mid \text{Vote}) = \begin{cases} .5 + \varepsilon_p & \text{if registered DEM}; \quad \varepsilon_p \in [0, .5) \\ .5 - \varepsilon_p & \text{if registered REP}; \\ .5 & \text{if registered UNA} \end{cases}$$

By construction, we assume that a voter is likely to vote for their party of affiliation, but not with certainty, and that an unaffiliated voter is equally likely to vote for either party. For simplicity and symmetry, we use the same $\varepsilon_p$ all party affiliations. $\varepsilon_p$ can be thought of as a measure of how strong we believe one's party affiliation actually is. We used the same method to calibrate the value of $\varepsilon_p$ as we did with $\varepsilon_v$. For our analysis, we chose $\varepsilon_p = .4$. While this interpretation is a bit extreme, choosing $\varepsilon_p = .4$ helped keep the results consistent the election outcomes.

The next step is to aggregate the data in such a way that it can be used to update our prior probabilities. The tendency of individuals to associate with others who are similar to themselves, known as homophily, is the idea that underpins our aggregating. The data shows that when people are broken down geographically into smaller groups they tend to have similar voting behavior and party preferences. Because empirical data is most readily available in a format broken down by municipalities, we choose this as our basis to split the data and aggregate over.[3] The aggregate predicted probability to vote in 2014 for

---

    subset of the full data set. The samples were chosen randomly to maintain the heterogeneity of the full data set and roughly keep the same proportion of voters from each municipality. We then calibrated $\varepsilon_v$ and $\varepsilon_p$ by tweaking their values until the average $p_{i,2014}(\text{Vote})$ and $p_{i,2014}(\text{Vote D} \mid \text{Vote})$ aligned with the past actual election results.

3  Note that if we had data broken down by zip, we could aggregate over zip instead of municipality.

voters living in municipality m is denoted $\bar{p}_{m,\,2014}(Vote)$ and the aggregate predicted probability to vote Democrat if a voter votes in 2014 for voters living in municipality m is denoted $\bar{p}_{m,\,2014}(Vote\ D\ |\ Vote)$. We calculate these as follows:

$$\bar{p}_{m,\,2014}(Vote) = \frac{\sum_{i \in m} p_{i,\,2014}(Vote)}{N_m}$$

$$\bar{p}_{m,\,2014}(Vote\ D\ |\ Vote) = \frac{\sum_{i \in m} p_{i,\,2014}(Vote\ D\ |\ Vote)}{N_m}$$

where $N_m$ is the number of voters in municipality m and the summation is over each of the voters in municipality m. In other words, these are the average probabilities for the voters in municipality m.

The final step in the model is to update the prior probabilities using the aggregate probabilities to obtain the posterior probabilities. Although our updating scheme is in the style of Bayesian inference, it is not a direct application of Bayes' rule. Instead, it is a principled updating scheme using arithmetic means. The update allows us to integrate the information about social consideration contained in our aggregate probabilities into our individual probabilities. We denote the posterior, updated likelihood to vote for individual i as $p'_{i,2014}(Vote)$ and the posterior, updated likelihood of a voter to vote democrat if he votes as $p'_{i,2014}(Vote\ D\ |\ Vote)$. We calculate these as follows:

$$p'_{i,2014}(Vote) = \tfrac{1}{2}\,(p_{i,2014}(Vote) + \bar{p}_{m,\,2014}(Vote))$$

$$p'_{i,2014}(Vote\ D\ |\ Vote) = \tfrac{1}{2}\,(p'_{i,2014}(Vote\ D\ |\ Vote) + \bar{p}_{m,\,2014}(Vote\ D\ |\ Vote))$$

where m is the municipality where voter i lives. The posterior probabilities are simply averages of the prior probability and the aggregate probability of the municipality where that voter lives. We chose to use the arithmetic mean of these two numbers so that the laws of probability hold. Specifically, by using the arithmetic mean, $p'_{i,2014}(Vote\ D\ |\ Vote)$ and $p'_{i,2014}(Vote\ R\ |\ Vote)$ must sum to 1, just as the priors did.

When we update p'$_{i,2014}$(Vote D | Vote) this enhances our estimation of which party a voter would vote for, if he votes. As the data shows, some municipalities have a very large percentage of unaffiliated voters. The way we define the priors reflects the fact that we have no information to indicate which way an unaffiliated voter will vote, so we simply assume it is equal for either party. This update is particularly useful to help discern which party the unaffiliated voters will likely align with, because it updates each individual voters probability with information derived from other voters in the same municipality.

## 4. Ranking Voters

Now that we have developed the model and applied it to our data to yield probabilities for each voter, we move on to the final step of ranking each of these voters. In our interpretation, we are not interested in a prediction of a voter's actual preferences; instead, they are relevant when looked at relative to each other. Ultimately, the ranking of voters is useful to a campaign because it gives a mathematical sound quantification of the individual voters that can will be most effectively targeted while campaigning.

Throughout this thesis, we have focused on calculating two quantities for a given voter: The likelihood of a given voter to actually vote, P$_{i,2014}$(Vote) and, if he votes, the likelihood for this voter to vote for a given party, P$_{i,2014}$(Vote D | Vote). The types of voters we are concerned with are support voters and swing voters, which can be broken down into turnout swing voters and persuasion swing voters. Support voters are those likely to support the party through activities such as donating, volunteering, or fund raising. These voters will near certainly turn out to vote and vote for the party they support; thus, they are not swing voters  We define a turnout swing voter as a voter who is highly probable to vote for the party using this model, but has a roughly average probability to turn out to vote. So we are confident this voter will vote for our party, we just want to try to convince him to actually turn out to vote. We define a persuasion swing voter as a voter who is highly probable to vote, but has a roughly average probability to vote for our party. This voter is likely to turn out and doesn't have a strong tendency to vote for either party; thus, we are hoping to convince him to vote for our party when he votes (Ghani, 2013). We will focus on turnout and persuasion voters, as these voters can be identified with the data set and model we are using.

Unlike presidential elections, congressional elections do not operate on an electoral college. That is, the winner of the election is determined solely by who has the greater number of votes over the entire district and not by who wins the most towns or precincts. Because of this, it makes sense that after updating each individual voter's probabilities with aggregate data from their town to pool all of this data together and try to rank each citizen on a district wide basis. This allows us to be able to rank each voter based on only their $p_{i,2014}$(vote) and $p_{i,2014}$(Vote D | Vote) and not have to consider a ranking system for the different towns in the district.

As mentioned above, we are interested in identifying two separate types of swing voters: turnout swing voters and persuasion swing voters. By definition, each of these voter types is determined by consideration of both of our probabilities. We implement a scheme using both $p_{i,2014}$(vote) and $p_{i,2014}$(Vote D | Vote). To do so, we define the following two quantities, which, by construction will attain their maximum value of zero when they most closely satisfy our criteria to be the type of swing voter we are interested in identifying:

$$r_{i,turnout} = -[p_i(\text{vote D | vote}) - (\overline{p(\text{vote D | vote})} + \varepsilon_{turnout})]^2 - [p_i(\text{vote}) - \overline{p(\text{vote})}]^2$$

$$r_{i,persuasion} = -[p_i(\text{vote D | vote}) - \overline{p(\text{vote D | vote})}]^2 - [p_i(\text{vote}) - (\overline{p(\text{vote})} + \varepsilon_{persuasion})]^2$$

Here, $r_{i,turnout}$ is the ranking position assigned to voter i to identify the most likely turnout swing voters and $r_{i,persuasion}$ is the position assigned to voter i to identify the most likely persuasion swing voter. The parameter values are as such: $\overline{p(\text{vote D} | \text{vote})}$ is the sample average of $p_i(\text{vote D} | \text{vote})$, $\overline{p(\text{vote D} | \text{vote})}$ is the sample average of $p_i(\text{vote D} | \text{vote})$, and $\varepsilon_{\text{turnout}}$ and $\varepsilon_{\text{persuasion}}$ are two positive constants which are used to shift their terms in the equation away from their respective averages towards higher likelihoods.

By construction, $r_{i,\text{turnout}}$ and $r_{i,\text{persuasion}}$ are non-positive functions that identiy, respectively, voters that best fit our definition of turnout and persuasion swing voters by having the value closest to the global maximum of this function, which is zero. Recall, for example, that for turnout swing voters we want to identify voters with an average chance of turning out and an above average chance of voting for a given party if they do turn out.

The first term should be closest to zero when the voter's likelihood to vote for a given party if they vote is above average. This gets slightly tricky as our function that updates our priors to our posteriors skews are probabilities in such a way that they are all drawn closer to the sample average. Particular attentions has to be paid to the data when choosing $\varepsilon_{\text{turnout}}$. Our intuition may tell us that a 75% chance of voting for a given party if they vote is above average. However, in our model, the maximum value of this probability is 77%, so the 75% actually implies near certainty they will vote for our party. This makes choosing $\varepsilon_{\text{turnout}}$ slightly tricky.

More specifically, the minimum and maximum of $p_i(\text{vote D} | \text{vote})$ closely correspond to the average value plus or minus only one standard deviation. Setting $\varepsilon_{\text{turnout}}$ to the one-third of the standard deviation allows the first term to achieve its value closest to zero when a voter's probability to vote for a given party if they vote is closest to the top two thirds of the sample. The second term is closest to zero when the voter's likelihood to turnout is closest to average, and decreases symmetrically as the voter's likelihood to turnout shifts from the average.

By virtue of our updating method, all of the individual's posterior probabilities are drawn closer to their municipality's average, so the idea of average and above average likelihoods are slightly skewed. In our sample, we see the posterior $P_{i,2014}$(Vote) range between 41% and 84% with an average of 55% and the posterior $P_{i,2014}$(Vote D | Vote) range between 49% and 77% with an average of 61%. Consequently, to identify swing voters we consider 55% to be average and 61% to be above average for $P_{i,2014}$(Vote) and 61% to be average and 67% to be above average for $P_{i,2014}$(Vote D | Vote).

Since each of our r values attains its maximum when it most closely identifies our voter of choice, we can simply sort each voter on either of these values independently to determine either the most persuadable voters to vote for a given party or the most persuadable voter to vote. This would be helpful in an ideal world where convincing a voter to vote and convincing a voter to vote for a given party were two fundamentally different processes. However, in reality, most campaign workers will not be proficient at either of these independently, but, instead, an unknown combination of the two. Thus, identifying the overall most persuadable voters will be most useful. Our final ranking is computed as follows:

$$r_{c,\ overall} = (1-c) * r_{turnout} + c * r_{persuasion} \quad\quad c \in [0,1]$$

This is a weighted average of the rankings of each type of swing voter. The weight, c, can be chosen to tend more towards turnout swing voters, persuasion swing voters, or an equal weighting of both. For the results below, we chose c = ½ for an equal weighting of both types of swing voters. Like our other ranking, the absolute most persuadable voter will have $r_{overall}$ = 0, thus a descending sort of voters by $r_{overall}$ gives our final ranking.

## 5. Results

Now that our model is complete, we can apply it to the data and systematically prescribe voting preferences for each voter to identify who the swing voters will likely be.

After ranking these voters, we have a clear idea of clusters of swing voters that could change the outcome of the election, if we effectively spend campaign resources trying to sway them. As Ghani (2013) highlights in his talk, we are aiming for results that are better than average, and we have achieved that. Our results don't claim to be accurate predictions of each voter's actual preferences; instead, a close enough approximation that will lead to useful results once compared to each other via ranking.

Recall that there are two types of swing voters to identify. The turnout voters who are very likely to vote for a given party and have an average likelihood to vote and the persuasion voters who are very likely to vote and have an average likelihood to vote for a given party. Also, recall that due to our updating scheme's tendency to draw the posterior probabilities closer to their average, we consider 55% average and 61% above average for $P_{i,2014}$(Vote) and 61% average and 67% above average for $P_{i,2014}$(Vote D | Vote).

Due to limitations of the excel spreadsheet used to implement the model, we show results for 5 municipalities in the district. Because we aggregate and update on a municipality-wide basis, we can divide the district into smaller subsets without the model breaking down. This is actually a robust feature of the model because we can adjust the model to one municipality, multiple municipalities, or all of the municipalities depending on our needs. We chose municipalities of varying population size to highlight that our rankings are not dependent on the number of registered voters. The specific municipalities chosen and their number of registered voters are: Bantam (551); Harwinton (4,406); Morris (1,665); Plainville (10,512); and, Waterbury (46,260).

The results presented below assume that we are trying to win votes for the Democratic party. The same analysis would apply if instead we were trying to win voters for the Republican party, so we do not present those findings. In our model, there are only a few different probability pairings for voters in a given municipality, so there will be many voters that share the same voting preferences in a given municipality. These groupings are a consequence of how we define are prior probabilities. [4] We have ranked the full set of voters, but group each of these clusters into bins of a given ranking and show one sample voter from each of these bins. The bin sizes vary, but are roughly proportional to the number of voters in the respective municipality. In all of the rankings below, the top 10 bins account for roughly 10% of the voters in our sample. Specifically, these bins represent roughly 6,000 voters from our sample of 65,000 voters. To help visualize the overlap between results in the different types of ranking, each color corresponds to a unique bin. Note that bins with no color appear only in that ranking scheme.

**Most probable turnout swing voters**

| VOTER_ID | PARTY | CITY | P(Vote 2014) | P(Vote D 2014 | Vote 2014) |
|---|---|---|---|---|
| 52345 | DEM | Harwinton | 0.54 | 0.69 |
| 23485 | DEM | Morris | 0.58 | 0.67 |
| 40525 | DEM | Bantam | 0.54 | 0.7 |
| 452 | DEM | Morris | 0.5 | 0.67 |
| 32405 | DEM | Plainville | 0.58 | 0.73 |
| 6723 | DEM | Harwinton | 0.63 | 0.69 |
| 30526 | DEM | Harwinton | 0.47 | 0.69 |
| 23252 | DEM | Plainville | 0.49 | 0.73 |
| 5624 | DEM | Bantam | 0.63 | 0.7 |
| 34362 | DEM | Bantam | 0.47 | 0.7 |
| 29602 | UNA | Waterbury | 0.59 | 0.57 |

**Most probable persuasion swing voters**

| VOTER_ID | PARTY | CITY | P(Vote 2014) | P(Vote D 2014 | Vote 2014) |
|---|---|---|---|---|
| 52134 | UNA | Waterbury | 0.59 | 0.57 |
| 9234 | DEM | Morris | 0.58 | 0.67 |
| 23452 | DEM | Harwinton | 0.63 | 0.69 |
| 65253 | UNA | Plainville | 0.58 | 0.53 |
| 42456 | DEM | Morris | 0.67 | 0.67 |
| 23452 | DEM | Bantam | 0.63 | 0.7 |
| 58492 | DEM | Harwinton | 0.54 | 0.69 |
| 3450 | UNA | Bantam | 0.63 | 0.5 |
| 17392 | DEM | Bantam | 0.54 | 0.7 |
| 53603 | UNA | Waterbury | 0.5 | 0.57 |

---

4 There are only 3 possible values for Pi,2014(Vote) and 4 possible values for Pi,2014(Vote D | Vote). This means there are only 12 unique pairs of prior voting probabilities for all voters. Our updating scheme shifts these values in a consistent way for each municipalities, so after being updated, each municipality will have 12 unique probability pairs. We are considering 5 different municipalities in our sample, and since each municipality has a different aggregate we have at most 60 unique posteriors probability pairings for all of our voters.

**Most probable overall swing voters**

| VOTER_ID | PARTY | CITY | P(Vote 2014) | P(Vote D 2014 \| Vote 2014) |
|---|---|---|---|---|
| 49303 | DEM | Morris | 0.58 | 0.67 |
| 14503 | DEM | Harwinton | 0.54 | 0.69 |
| 62810 | UNA | Waterbury | 0.59 | 0.57 |
| 23025 | DEM | Bantam | 0.54 | 0.7 |
| 47289 | DEM | Harwinton | 0.63 | 0.69 |
| 9834 | DEM | Bantam | 0.63 | 0.7 |
| 1293 | DEM | Morris | 0.5 | 0.67 |
| 21203 | DEM | Plainville | 0.58 | 0.73 |
| 34928 | DEM | Morris | 0.67 | 0.67 |
| 32538 | UNA | Waterbury | 0.5 | 0.57 |
| 52352 | UNA | Plainville | 0.58 | 0.53 |

These results show that there is a fair amount of overlap between the first and second ranking scheme, and the results from the third ranking scheme are almost entirely of entries from the first two. The results also agree with our intuition that most of the turnout swing voters are declared Democrats while a larger number of persuasion swing voters are Unaffiliated. This shows that our model helps to identify the true voting preferences of an unaffiliated voter, satisfying one of the main objectives of our model.

With these results, we can now micro-target voters and know where to focus our campaign efforts to try to sway the swing vote in our direction and achieve our overall goal of winning the election. The results shown here account for roughly 6,000 voters (which is roughly 10%) in our district. This provides enough voters to focus on such that we can attempt to influence the election, but not so many that the effort becomes intangible. Beyond that, the power of the ranking system is that the top of the list identifies the most significant swing voters, yet the more resources a campaign has, the farther down the list of voters they can go.

**6. Theory**

This model is not an implementation of any single theory or abstract model, but is instead inspired from multiple different theoretical frameworks found in economic, political, and mathematical literature. Rational choice theory and statistical mechanics ideas were used to develop the theory behind the model and the framework our voters interact within, and statistics and Bayesian inference were used to obtain useful results from our data within this constructed framework.

Classical rational choice theory posits that all socioeconomic agents act rationally in order to maximize their utility in regards to their own self interest. In their treatment on social interactions, Christakis and Fowler (2009) highlight that when you apply rational choice theory to political science, due to the assumption of rationality, it is very unlikely for a voter to vote because the costs associated with voting are too high relative to the payoffs. Fowler (2006) explains that since each vote is less and less likely to impact an election, the cost to vote strongly outweighs any payoff to the voter derived from affecting the outcome of an election; therefore, a low voter turnout is predicted. Empirical results contradict theory and show that voter turnout is generally very high. In his Nobel Prize winning paper, McFadden (2001) accounts for disparities between theoretical predictions and empirical findings by showing that people do not always act as completely rational representative agents, but instead, perceptions of others influence an agent's economic decisions.

When local incentive structures and social interaction are more explicitly considered, rational choice theory much more effectively explains social phenomena. These approaches factor in interaction among agents by incorporating the utility of nearby agents into a given agent's own utility function. Now, a voter's utility is not simply a maximization problem over his own self interest, but incorporates the utility of his neighbors. Smith, LaGatta, and Bueno de Mesquita (2013) demonstrate an interesting way to incorporate local incentive structures in voting models. They suppose that the agents can win a prize for their district if the candidate they vote for wins. Their consideration of local incentives for their community motivates them to turn out and vote, and this help explain a large voter turnout.

Social interaction is playing a larger and larger role in developing economic theory. The interaction of agents on an individual scale affects the aggregate state of a system. In our model, the actions and interactions of voters on an individual scale have a discernible effect on the winner of an election, and this agrees with our intuition. Statistical mechanics was developed to describe and draw a link between the microscopic and macroscopic states of large physical systems and to explain aggregate phenomenon, such as magnetization and changes in state of matter. Durlauf (1999) states: "Just as statistical mechanics models explain how a collection of atoms can exhibit the correlated behavior necessary to produce a magnet, social science models wish to explain interdependent behaviors. The basic idea in statistical mechanics—that the behavior of one atom is influenced by the behavior of other atoms—is thus similar to the social science claim that one individual's decisions depend upon the decisions of others; therein lies the possibility of a common mathematical structure." This analogy applies directly to our model when considering interdependent behavior among voters. Much more importantly, he shows in this paper that the mathematical constructs underlying statistical mechanics do indeed hold true in an economic context.

Looking at economics through the lens of statistical mechanics helps to break down some of the flaws in the assumption of a representative agent and allows us to account for heterogeneous preferences in utility maximization problems. Aoki and Yoshikawa (2011) present their interpretation of macroeconomics with an emphasize on interaction among agents and rely on the mathematical tools that statistical mechanics has established, which helps set the context for the use of these models. Bouchaud (2011) presents few specific examples of how these mathematics apply to binary choice models. This applies to our model as our voters are faced with a sequence of two binary choice: 1) Vote or Abstain; 2) If they vote, vote democrat or vote republican.

In our model, when updating each individual's probabilities with aggregate data from the municipality, we implicitly assume that nearby people will have similar preferences. This assumption lies at the core of our model and it is the method by which we integrate social interaction into each individual's preferences. Contucci and Ghirlanda (2007) show that grouping of voters with similar preferences is a consequence of the statistical mechanics framework and justifies our updating with municipality-wide aggregates.

Most of the municipalities we considered have a large number of citizens, so there seems to be a natural fit to the large system nature of statistical mechanics. However, there are a few smaller municipalities that we work with and our intuition still holds here. Lee, Broedersz, and Bialek (2013) shows that statistical mechanics can be used to explain the behavior of the Justices of the Supreme Court of the United States, even though there is a very small number of agents in this system, namely, 9. This paper also presents a very nice small scale example of determining voter preferences in a context where, due to strong party divisions in the Supreme Court, swing voters can also make a large impact on the final decision of the court.

A more technical exploration of the mathematical components of statistical mechanics that apply to economics can be seen in Brock and Durlauf 's (2002) coverage of social interaction in economic behavior. They demonstrate, "models which incorporate terms reflecting the desire of individuals to conform to the behavior of others in an environment of non cooperative decision making." Specifically, they show how the mean field version of the Ising model has the same underlying probability structure as their economic model that describes the collective behavior of the interacting economic agents. We refer to Fedele (2011) to enhance our intuition of the statistical mechanic theory involved in mean field models and applications to collective behavior of interacting particles.

Rational choice theory and statistical mechanics applications to economics provide a useful framework for explaining the theory behind the model, but we need tools to analyze the data and implement the numerical portion of the model. Basic probability and statistics allow us to make a first order guess at reasonable probabilities for a given voter, but it is Bayes' rule that gives a mathematical sound method for updating probabilities based on observed information and intuition. Formally stated, Bayes' rule says:

$$P(A|B) = P(B|A)*P(A) / P(B)$$

Here, P(A) is our initial belief, the prior probability, P(A|B) is our belief having accounted for B, the posterior probability, and P(B|A)/P(B) is the support A provides for B, our updating function (Lee, 2012). In use, this allows one to compute the probability of an event to occur with the knowledge that some other event has already been observed.

A more applicable interpretation of Bayes' Rule is present by Lee (2012). When properly defined probability distributions exist, Bayesian statistics state that the posterior distribution is proportional the prior distribution times an update function:

$$\text{Posterior Distribution} \propto \text{Prior Distribution} \times \text{Update Function}$$

Bayes' rule allows us to change our belief of the likelihood of an event based on information we have learned. It is a way to formally revise a probability when new observations are made.

Our model relies heavily on the idea of Bayes' rule, but is not an explicit application of it. We do not define a proper probability distribution for each voter in our sample, so we cannot apply Bayes' rule in its literal format. Instead, we use the idea behind Bayes' rule that we can update our probability predictions by adding more information. Bolstad (2007) and Kruschke (2010) provide many examples of Bayesian inference that help show the power of this tool. We apply this in our model by establishing our priors in a way that seems reasonable. We then, aggregate this information on a municipality-wide basis to integrate the social considerations discussed early. Using a a principled approach, we update our prior probabilities by averaging them with the aggregate data to arrive at our posterior probabilities.

The averaging scheme employed is where we violate Bayes' rule in its strict sense, since it does not preserve the underlying probabilities. This means our posteriors are no longer an absolute expression of our voters preference. The power of our model is that this does not matter. Because we are using a ranking scheme to identify the swing voters, only the probabilities relative to one another are significant, and our method of updating preserves this while still integrating more information into these predictions.

## 7. Conclusion

In the end, we were able to develop a model that takes a single data set containing information on all of the voters in a given district. We were able to use our model to make predictions on voting probabilities of these voters, then use a Bayesian themed updating to make these predictions more accurate by adding further information to them. After applying our ranking scheme to the individual voters, we are left with an ordering of all of the voters in the district that will be useful in a very practical sense for a political campaign that is trying to identify and sway swing voters in the hopes of winning the election.

The model draws insight from multiple ideas and integrates them in a reasonable way. Our model does not give realistic probabilistic for each voter. It does make enough headway in updating one's likelihood to vote, and importantly, it draws on aggregate data to address the important issue of which way swing voters will go.

Building a good model is a very cyclic process. Only after gaining the intuition from completing the model can one look back and have a clear idea where the model could be improved. Now that model is done and the work has been thought through, there is an excellent opportunity to consider a few areas of improvement that would yield the most substantial improvement in the model.

One of the very nice aspects of our model is that we used only one data set throughout the project to compute all of our probabilities. We gleaned useful results from this data by both using it directly to calculate further information and reexamine the data from a different perspective when we aggregated the numbers. An obvious weakness is the data set contains only information from a fixed point in time. Data is available for the number of voters registered for each election and the turnout for each year. Incorporating this data will help address this issue and will better allow us to see how voter turnout fluctuates year to year and make a better real world prediction for 2014 voter turn out. That is, our numbers could be useful on an absolute scale, not just relative to each other.

Another short coming of our model is that it does not take into account past election results. Specifically we do not incorporate any information on the actual vote turnout for a given party to help determine which way unaffiliated voters will tend to vote. Our model now implicitly assumes that the unaffiliated voters in a town are likely going to align with the party to which the majority of a municipality's voters are registered members of. This agrees with the theoretical underpinning our model, in that voter's located near each other will tend to affect each others' decisions and it could be concluded that they will likely vote the same way. However, in many cases, a majority of voters are unaffiliated, so the assumption that a small number of voters determines the preferences of a larger percentage of voters may be a flawed one.

Take, for example, Waterbury where 49% of voters are registered Democrat, 12% registered Republican, and 39% are unaffiliated. Our model would indicate that unaffiliated voters would vote Democrat. However, looking to average actual election results over the last 10 years show a roughly 59% Democrat to 41% Republican split. These results contradict our model and suggests that most unaffiliated voters actually vote Republican. However, only through a more careful application of Bayes' Rule can we attempt to better discern which way unaffiliated voters will actually vote. To do so properly, we need to use the following quantities which are known from our model: 1) The likelihood of being each type of voter, 2) The likelihood of each type of voter voting for a given party if they vote, and, 3) The likelihood to actually vote for each of these types of voters. Importantly, the last quantity needed is the actual percentage of voters that vote Democrat and Republican, and establishing this quantity will necessitate incorporate additional data. Once found, we can use these values to update our priors in a strictly Bayesian methodology to obtain posteriors that more accurately reflect an unaffiliated voter's true voting preferences.

Aside from trying to better update our priors through incorporation of further data, the model can be improved by considering further ranking schemes. After all, this is our method for determining the voters to spend campaign resources on. Once these other rankings are achieved, it would be interesting to look at how similarly they rank each voter. A final rank could be constructed as an aggregate of all of individual rankings.